# SIMULATIONS OF X-RAY CLUSTERS

Julio F. Navarro[1], Carlos S. Frenk[1], and Simon D.M. White[2]

[1] *Physics Department, University of Durham, Durham DH1 3LE, England*
[2] *Institute of Astronomy, Madingley Road, Cambridge CB3 0HA, England*

**ABSTRACT**

We present simulations of the formation and evolution of galaxy clusters in the Cold Dark Matter cosmogony. Clusters with a wide range of mass were selected from previous N-body models, and were resimulated at higher resolution using a combined N-body/Smooth Particle Hydrodynamics code. The effects of radiative cooling on the gas are neglected. While many present-day clusters are predicted to be undergoing mergers, the density profiles of those that are approximately in equilibrium are all very similar, both for the gas and for the dark matter. These profiles show no sign of a uniform density core and steepen gradually from the centre outwards. The standard $\beta$-model is a reasonable fit over most of the radius range observable in real clusters. However, the value obtained for the slope parameter $\beta_f$ increases with the outermost radius of the fit. Temperature profiles of different simulated clusters are also similar. Typically the temperature is almost uniform in the regions which emit most of the X-ray flux but drops at larger radii. The gas temperature and dark matter velocity dispersion in equilibrium clusters give values of $\beta_T \equiv \mu m_p \sigma_{DM}^2/kT$ which are consistent with unity provided an X-ray emission-weighted temperature is used. Larger values of $\beta_T$ are found in merging objects where there is a transient boost in the velocity dispersion of the system. Thus $\beta_T > 1$ may be an observational indicator of merging in real clusters. The similar structure of clusters of differing mass results in scaling relations between the X-ray and dynamical properties of clusters identified at any given redshift. These scalings are inconsistent with the observed slope of the luminosity-temperature relation or the observed sense of evolution of the cluster luminosity function. This suggests that the central properties of the intracluster medium are determined by non-gravitational processes such as radiative cooling or substantial pre-heating at high redshift.

## 1 INTRODUCTION

The dynamical timescales characteristic of galaxy clusters are a large fraction of the age of the Universe. As a result, cluster formation is an ongoing process which can be observed directly in nearby systems, and many properties of the initial conditions from which clusters formed are still visible in their present structure and dynamical state. Detailed theoretical study of cluster evolution requires modelling the dynamically dominant dark matter, the hot X-ray emitting intracluster gas and the population of cluster galaxies. Although the behaviour of these three components is strongly coupled, theoretical studies of galaxy clusters have progressed by stages, focussing in the first instance on the evolution of the dark matter and on the dynamics of an unevolving galaxy population. Only in the past few years has detailed modelling of the dynamics of the intracluster gas become possible.

The simplest model which might capture the essential gas dynamics of cluster formation is one in which a nonradiative ideal gas moves in the evolving gravitational field of the dissipationless dark matter. This approximation may miss some essential physics; although the radiative cooling time exceeds the Hubble time for most of the gas in observed clusters, much of their X-ray emission often comes from a central region where the opposite inequality holds (Fabian *et al.* 1991). In addition, heating and cooling within the smaller but denser clumps which were the precursors of observed clusters could have substantially altered the thermodynamic state of the gas. Nevertheless, it is clear that a detailed understanding of the simple problem is a prerequisite for studying more realistic (but inevitably more complex) situations. This problem also provides a suitable testbed for assessing the reliability of the simulation techniques which are increasingly being used to investigate cluster evolution.

A crude but useful analytic model for cluster formation assumes spherical symmetry and follows accretion onto an initially overdense perturbation. Each shell expands to a maximum radius, turns around and falls back onto the perturbation. Shells of dark matter pass through the centre and then oscillate within the nonlinear body of the cluster, while infalling gas shells encounter an outwardly moving accretion shock which thermalizes their kinetic energy. Inside this shock the gas is approximately in hydrostatic equilib-rium. Bertschinger (1985) gives a similarity solution for the density and temperature structure of the two components in the particular version of this infall problem which was originally set out by Gunn and Gott (1972).

If the structure of all clusters is similar, as is expected in the above model, then simple scaling relations can be derived for the characteristic properties of clusters which formed from a scale-free spectrum of initial density fluctuations (*e.g.* White 1982). Each cluster present at redshift $z$ can be characterised by the mass $M$ contained within a sphere of some given overdensity. The mean temperature, $T$, of the cluster and its X-ray luminosity, $L_X \propto M \rho T^{1/2}$ ($\rho \propto (1+z)^3$ is a characteristic density), then scale as

$$T(M, z) \propto M^{2/3} (1+z) \qquad (1)$$

and

$$L_X (M, z) \propto M^{4/3} (1+z)^{7/2}. \qquad (2)$$

Kaiser (1986) used these relations to make detailed predictions for the evolution of the cluster population.

More recent data have shown that these scaling relations fail in a number of important respects. For example they imply $L_X \propto T^2$ which is flatter than the observed relation, $L_X \propto T^{2.7 \to 3.5}$ (Edge and Stewart 1991a,b, Henry and Arnaud 1991). Furthermore, at fixed temperature or mass the X-ray luminosity is predicted to be larger at earlier times, giving rise, for realistic cluster mass functions, to an increase in the comoving number density of bright clusters with redshift (Kaiser 1991, Blanchard *et al.* 1992, Bartlett and Silk 1993). Current data imply just the opposite trend (Edge *et al.* 1990, Gioia *et al.* 1990, Henry *et al.* 1992, Castander *et al.* 1994, Bower *et al.* 1994). As Kaiser (1991) and Evrard and Henry (1991) pointed out, better results are obtained if the intracluster gas is assumed to have been heated to high entropy prior to cluster collapse (perhaps by energy injected during galaxy formation) and to have evolved adiabatically thereafter. If the central regions of all clusters have the same entropy, then eqn (2) is replaced by

$$L_X (M, z) \propto M^{11/6} (1+z)^{11/4}, \qquad (3)$$

(Evrard and Henry 1991), so that $L_X \propto T^{11/4}$, independent of redshift.

In recent years several studies have used N-body/hydrodynamic codes (both Lagrangian and Eulerian) to simulate the evolution of the intracluster medium in the nonradiative limit (Evrard 1990a,b; Thomas and Couchman 1992, Kang *et al.* 1994, Bryan *et al.* 1994). This work relaxed the assumption of spherical symmetry and assumed instead that clusters form from a Gaussian random field of initial density fluctuations. In some cases, the final configurations look remarkably like real clusters: the temperature distribution is approximately isothermal throughout the body of the cluster and the gas density is well fit by the "β-model" often used to characterize observed clusters. However, not all simulations give the same results. For example, Thomas and Couchman's cluster has a steeper temperature gradient and a significantly different density profile than those in Evrard's simulations. In addition, there is no consensus on the basic issue of whether the simulations obey the expected scaling relations. For example, the total X-ray luminosity of the rich cluster simulated by Evrard (1990b) seems to increase with time, while the cluster population in the simulations of Kang *et al.* (1994) and Bryan *et al.* (1994) deviates dramatically from the scaling behaviour predicted in eqns (1) and (2).

In this paper we present new N-body/hydrodynamic simulations designed specifically to explore the simple nonradiative model, including the case where the gas is initially preheated. Our aims are twofold. We wish to test the applicability of the scaling laws in models with realistic cosmological initial conditions and realistic hydrodynamics. We also wish to use the relatively simple behaviour expected theoretically in this case to assess the reliability and the limitations of our numerical techniques. Our initial conditions assume a universe dominated by cold dark matter, the paradigm of hierarchical clustering theories. We focus primarily on the evolution of the distributions of gas and dark matter within our clusters. We also examine in some detail the well-known "β-discrepancy" in clusters, and we assess the reliability of cluster mass estimates based on X-ray data. Our simulations are similar to but have better spatial resolution than those presented by Evrard (1990a,b) and Thomas and Couchman (1992).

The rest of this paper is organized as follows. In Section 2 we describe our numerical methods and the parameters of our models. In Section 3 we discuss the evolution and the internal structure of our clusters, we compare them with Bertschinger's (1985) spherical accretion model, and we check how well they obey a similarity solution. In Section 4 we compare our models with real clusters and we discuss the β-discrepancy. In Section 5 we describe simulations with pre-heated gas. Finally, we summarize our results in Section 6.

## 2   THE SIMULATIONS

### 2.1   The Code

Our code is designed to follow the evolution of a mixture of collisional and collisionless fluids in three dimensions. It is based on the Smooth Particle Hydrodynamics (SPH) technique and uses a nearest-neighbour binary tree to compute gravitational forces. It is fully Lagrangian, free from symmetry restrictions, and has a large dynamic range. It is thus well suited to the study of irregular systems with structure on a wide range of scales.

SPH is a Monte Carlo technique which represents a gas by "particles" which are used as interpolation centres for the computation of thermodynamic quantities and their gradients. Each particle is assigned a mass, a density, and a temperature; pressure is then obtained from an equation of state and gradients are estimated by convolving the particle distribution with appropriate kernel functions. In this paper we assume an ideal gas, $p = (\gamma - 1)\rho u$, where $p$, $u$, and $\rho$ denote pressure, specific internal energy, and density respectively; we take $\gamma = 5/3$. Artificial viscosity terms in the equations of motion allow a crude but energetically consistent representation of shocks. In our code gravitational forces between particles are softened using a spline kernel of the same form as the interpolation kernel, although the corresponding softening length is set independently and is kept constant in time. Collisionless (dark matter) particles move under the influence of gravity alone. Excellent reviews of SPH can be found in Benz (1990) and Monaghan (1992), a description of the binary tree we use is in Benz *et al.* (1990), our particular implementation of SPH and several tests of our code are described by Navarro and White (1993), and technical descriptions of SPH as applied to cosmological problems are given by Evrard (1988) and Hernquist and Katz (1989).

## 2.2 Initial Conditions

The clusters we have studied were selected from the cosmological N-body simulations carried out by Frenk *et al.* (1990). These simulations modelled cubic regions, 360 Mpc on a side, within an $\Omega = 1$, $h = 0.5$ CDM universe. (All physical quantities quoted here assume $h = 0.5$.) Six dark matter clumps with one-dimensional velocity dispersions ranging from 400 to 1300 km/s were chosen at an output time when the rms linear overdensity in a sphere of radius 16 Mpc took the value, $\sigma_8 = 0.63$. The particles in each of these clumps were traced back to the initial conditions $(1 + z_i = 4.74)$ and a cubic region containing all of them was defined. This region was then filled with 10648 new particles on a cubic grid which were perturbed using the original waves of the Frenk *et al.* initial conditions, plus additional waves chosen to represent the CDM power spectrum between the original resolution limit and the Nyquist frequency of the new particle grid. To ensure that each cluster in our new simulations is represented by approximately the same number of particles, we chose the size of this "high-resolution box", $L$, as indicated in Table 1. The distribution of mass outside the high-resolution region was modelled by $\sim 6000$ particles of radially increasing mass using a technique similar to that described by Katz and White (1993). Overall, this procedure extends the dynamic range in mass of the original simulation by a factor of $\sim 20$ for the most massive cluster and $\sim 660$ for the least massive cluster. Evolving our initial conditions with an N-body code we are able to recover the clumps in the original low resolution simulations, but new small-scale substructure, arising from the additional high frequency waves, is sometimes noticeable.

Within our high resolution region we placed a gas particle on top of each of the dark matter particles and gave it a very low temperature ($\sim 15\,K$). This ensures that gas dynamical effects do not become important until the first nonlinear structures form. The mass of a gas particle, $m_{gas}$, is determined by the baryon fraction $\Omega_b$ which we took to be 0.1. (Note that since cooling is not included, our results can, to a good approximation, be directly scaled to other values of $\Omega_b$.) The gravitational softening, $h_g$, was set equal to 100 kpc in all cases. Numerical parameters for all our simulations are listed in Table 1. The symbols will be defined in the text when they are first mentioned.

In addition to our six standard runs, labelled $CL1$-6, we ran three further simulations in which the gas was "preheated" at $z = 3$. This was accomplished by raising the temperature of each gas particle so as to give a uniform specific entropy, $s = \ln(T/\rho^{2/3})$, equal to the specific entropy at the centre of $CL1$, our most massive cluster, at $z = 0$. Our three "preheated" models are labelled $EP1$ to $EP3$ in Table 1 and have the same initial conditions as the standard models $CL1$, $CL3$, and $CL5$ respectively. Simulations $CL1 \times 3$ and $CL1/2$ use the same initial conditions as that of run $CL1$, but the total number of particles has been increased by a factor of $\sim 3$ and decreased by half, respectively.

## 2.3 Estimates of X-ray luminosity and code units

The *bolometric* X-ray luminosity of a cluster is given by the volume integral of the emissivity,

$$L_X = \int_V L(T)dV, \tag{4}$$

where $L(T) = n^2 \Lambda(T)$. Here, the gas number density, $n = \rho/\mu m_p$, where $m_p$ is the proton mass, $\mu = 0.6$ for a fully ionized primordial plasma (Spitzer 1978), and $\Lambda(T)$ is the cooling function which, at the temperatures relevant to rich clusters, can be well approximated by

$$\Lambda(T) = 1.2 \times 10^{-24} \left( \frac{T}{1\,\mathrm{keV}} \right)^{1/2} \mathrm{erg\,cm^3\,s^{-1}}, \tag{5}$$

For the simulations we use the estimator,

$$L_X = 1.2 \times 10^{-24} (\mu m_p)^{-2} m_g \sum_{i=1}^{N_{gas}} \rho_i T_i^{1/2} \, \mathrm{erg\,s^{-1}}, \tag{6}$$

where $m_g$ is the mass of a gas particle, and $\rho_i$ and $T_i$ are the density and temperature (in keV) at the position of the $i$-th gas particle.

The computational units of the code are given by:

$$\mathrm{G} = 1$$

$$[\mathrm{mass}] = 10^{10}\, m_u\, \mathrm{M_\odot}$$

$$[\mathrm{distance}] = 1\, d_u\, \mathrm{kpc}$$

$$[\mathrm{time}] = 4.71 \times 10^6 \left( \frac{d_u^3}{m_u} \right)^{1/2} \mathrm{yrs}$$

$$[\mathrm{velocity}] = 207.4 \left( \frac{m_u}{d_u} \right)^{1/2} \mathrm{km/s}$$

$$[\mathrm{density}] = 6.77 \times 10^{-22} \left( \frac{m_u}{d_u^3} \right) \mathrm{g/cm^3}$$

$$[\mathrm{energy/mass}] = 4.3 \times 10^{14} \left( \frac{m_u}{d_u} \right) \mathrm{erg/g}$$

$$[\mathrm{energy/time}] = 5.75 \times 10^{43} \left( \frac{m_u}{d_u} \right)^{5/2} \mathrm{erg/s}.$$

Physical quantities from the simulations can be consistently rescaled to other situations by changing the values of $m_u$ and $d_u$. The numbers quoted throughout this paper correspond to $m_u = d_u = 1$.

## 3 RESULTS

### 3.1 Evolution

Figure 1 shows the evolution of the distributions of dark matter and of gas in $CL1$, our most massive cluster. Overall, the two distributions look very similar although, as discussed below, there are significant differences of detail. The cluster grows primarily through mergers, as predicted by the analytic model originally due to Press and Schechter (1974), and extended by Bond *et al.* (1991) and Bower (1991). At $z = 0.84$ and 0.40, the fraction of the present mass of the cluster that is in clumps (identified using a friends-of-friends algorithm) of more than $\sim 100$ particles is 0.48 and 0.75, respectively, compared with 0.37 and 0.68 predicted by the analytic Press-Schechter theory. Increasing the number of particles of the simulation by a factor of three (run $CL1 \times 3$) brings these numbers down to 0.45 and 0.73, improving the agreement with the analytic model. At each stage gas collapses into clumps with the dark matter and is shock-heated to the relevant virial temperature. This hierarchical growth continues to the present and results in significant irregularity in our final clusters. Figure 2 shows the final configuration of the gas in clusters $CL2$-$CL6$. In several cases, substructure is still clearly discernible and some of the clusters have massive companions nearby.

**Table 1.** Cluster properties ($z = 0$).

| Model | $L$ [Mpc] | $M_{200}$ [$10^{10} M_\odot$] | $m_{gas}$ [$10^{10} M_\odot$] | $r_{200}$ [Mpc] | $\sigma_{DM}$ [km/s] | $\Omega_{gas}$ | $R_{gas}/R_{DM}$ | $T_X$ [K] | $L_X$ [erg/s] |
|---|---|---|---|---|---|---|---|---|---|
| $CL1$ | 45.0 | 2.29e5 | 5.94 | 3.43 | 1.17e3 | 0.093 | 1.17 | 1.10e8 | 3.98e45 |
| $CL2$ | 38.2 | 2.19e5 | 3.58 | 3.36 | 1.15e3 | 0.087 | 1.11 | 8.13e7 | 2.34e45 |
| $CL3$ | 31.0 | 1.05e5 | 1.94 | 2.63 | 8.91e2 | 0.086 | 1.11 | 4.57e7 | 1.02e45 |
| $CL4$ | 25.9 | 5.13e4 | 1.14 | 2.08 | 7.24e2 | 0.087 | 1.21 | 3.47e7 | 5.75e44 |
| $CL5$ | 20.9 | 1.15e4 | 0.59 | 1.26 | 4.37e2 | 0.091 | 1.09 | 1.26e7 | 6.46e43 |
| $CL6$ | 14.0 | 9.77e3 | 0.18 | 1.19 | 3.72e2 | 0.089 | 1.22 | 1.00e7 | 4.07e43 |
| $CL1 \times 3$ | 45.0 | 2.34e5 | 1.93 | 3.44 | 1.17e3 | 0.097 | 1.21 | 1.11e8 | 4.36e45 |
| $CL1/2$ | 45.0 | 2.43e5 | 11.9 | 3.48 | 1.18e3 | 0.094 | 1.33 | 1.15e8 | 2.88e45 |
| $EP1$ | 45.0 | 2.29e5 | 5.94 | 3.43 | 1.17e3 | 0.094 | 1.19 | 1.17e8 | 3.09e45 |
| $EP2$ | 31.0 | 1.05e5 | 1.94 | 2.63 | 8.91e2 | 0.085 | 1.17 | 6.02e7 | 6.16e44 |
| $EP3$ | 20.9 | 1.12e4 | 0.59 | 1.26 | 4.27e2 | 0.071 | 1.67 | 1.86e7 | 1.51e43 |

The rate at which mass is added to the clusters is illustrated in Figure 3. We plot the evolution of the dark mass within a sphere of current mean overdensity of 200. At each time we centre this sphere on the most massive clump present. The curves are normalized to the mass $M_{200}$ contained at the final time, and different symbols correspond to different simulations. Mergers stand out in this plot as sudden jumps in mass. As expected in an $\Omega = 1$ universe, the formation of rich clusters is a very recent phenomenon. Half the clusters in Figure 3 have doubled their mass since $z = 0.2$, and most have undergone substantial mergers since that time. This explains the substantial irregularity seen in Fig. 2. As pointed out by Forman and Jones (1990), a large fraction of real X-ray clusters show significant substructure. Several authors (eg Richstone *et al.* 1992, Lacey and Cole 1993, Kauffmann and White 1993, Evrard *et al.* 1994b) have noted that this argues in favour of a high value of $\Omega_0$.

Figure 4 shows the evolution of the thermodynamic properties and the X-ray luminosity of the intracluster gas. The gas which ends up near the median radius of a cluster typically reaches high density as the first resolved structures form; its average density remains essentially constant thereafter (dotted lines). The same is almost true for the gas which ends up in a cluster core; its mean density increases only slightly with time after the initial phase of evolution (solid lines). The mean density of the gas which ends up in the outer regions actually declines slowly over most of the evolution (dashed lines). The mean temperature of the gas which ends up in each of the three regions shows a slow but steady increase due to adiabatic compression and shock-heating. The shocks caused by accretion and mergers tend to be mild because the relative speeds are generally not highly supersonic. This is particularly so in the central regions where the entropy increases only slowly after the collapse of the first subclumps. In the outer parts shocks due to recent mergers are more effective in raising the entropy. The X-ray luminosity is determined primarily by the gas density near the core and depends only weakly on temperature (eqs 4-6). On average it changes little after $z = 1$, but it fluctuates whenever a merger drives oscillations in the central density. (These can have amplitudes of up to a factor $\sim 3$.) After a merger the X-ray luminosity is generally slightly higher than the combined luminosity of the progenitors.

The results shown in Figure 4 are similar to those of previous simulations (see e.g. Figure 1 of Thomas and Couchman 1992 and Figure 4 of Evrard 1990b). Notice, however, that they do not seem consistent with the predictions of the simple scaling laws we gave in the Introduction. For example, these predict the luminosities of clusters to decrease with time. We now investigate in detail the internal structure of our clusters and compare their properties and evolution with the predictions of the spherical infall model and the self-similar scaling relations.

### 3.2 Density profiles

At the present epoch several of our clusters have substantial substructure within the virial radius which reflects recent mergers (see Figure 2). In the present section we wish to study the "equilibrium" state of clusters. Hence for those clusters with significant substructure at $z = 0$ we consider slightly earlier epochs when the most recently accreted massive clump was still outside the virial radius. In all cases we can choose such a time at $z \lesssim 0.2$. The results presented in this and the following two subsections refer to these "equilibrium" configurations. It is important to note that they *cannot* be considered typical of nearby clusters. Rather they are typical of nearby *regular* clusters, which may be a minority of all clusters.

**Table 2.** Density profile fits.

| Model | $\Delta_0$ [gas] | $r_c^{(a)}$ [gas] | $\beta_f$ [gas] | $\Gamma$ [gas] | $\Delta_0$ [DM] | $r_c^{(a)}$ [DM] | $\beta_f$ [DM] |
|-------|------|------|------|------|------|------|------|
| $CL1$ | 1.57e3 | 0.106 | 0.86 | 1.30 | 2.40e4 | 0.082 | 0.86 |
| $CL2$ | 2.77e3 | 0.101 | 0.77 | 1.40 | 9.17e4 | 0.054 | 0.75 |
| $CL3$ | 4.22e3 | 0.101 | 0.90 | 1.38 | 2.91e4 | 0.125 | 1.05 |
| $CL4$ | 1.89e3 | 0.074 | 0.74 | 1.35 | 3.20e4 | 0.066 | 0.78 |
| $CL5$ | 1.28e3 | 0.127 | 0.84 | 1.15 | 2.42e4 | 0.163 | 0.90 |
| $CL6$ | 1.31e3 | 0.098 | 0.80 | 1.42 | 1.76e4 | 0.093 | 0.85 |

$^{(a)}$ $r_c$ is in units of $r_{200}$. All fits are within $r_{200}$.

Density profiles averaged over spherical shells are shown for the gas and the dark matter in Figures 5a and 5b respectively. The center of each system was defined by an iterative procedure that computes the center of mass inside a spherical region and then shrinks that spherical region until about 100 particles are left. This picks up the center of the most massive clump in the volume of interest. The two components have similar profiles which are poorly described by a single power-law. Their logarithmic slope steepens from $\sim -1$ near the centre to $\sim -3$ at large radius. There is no evidence for a constant density core outside the region where the gravitational softening is important. (At twice the softening radius the density in all clusters has dropped by at least a factor of two from its value at $h g$.) Scaled density profiles for the six clusters are shown in Figure 6a. Here we plot overdensity (density divided by the mean density of the universe at the time when the cluster is identified) as a function of the dimensionless radial variable, $r/r_{200}$, where $r_{200}$ is the radius of the sphere within which the mean interior overdensity is 200. The scaled profiles are all remarkably similar outside the region affected by gravitational softening. Bumps in the profiles beyond $r_{200}$ are due to companion systems which are falling onto the cluster. Despite the lack of a constant density core, the scaled density profiles of both gas and dark matter are fitted moderately well by the "$\beta$-model",

$$\rho/\overline{\rho} = \Delta_0 \left(1 + (r/r_c)^2\right)^{-3\beta_f/2}, \qquad (7)$$

over all but the innermost regions. This fit is shown as the thick solid line in Figure 6b which has $\beta_f = 0.8$, $\Delta_0 = 1.4e3$ and $r_c/r_{200} = 0.1$ for the gas and $\beta_f = 0.8$, $\Delta_0 = 2.2e4$ and $r_c/r_{200} = 0.09$ for the dark matter. The best-fit parameter values for the individual clusters are listed in Table 2. (Note that while the slope of the outer profiles is quite well determined, the values of $r_c$ and $\Delta_0$ are affected by poor resolution near the centre and have highly correlated uncertainties.)

While the $\beta$-model has traditionally been used to describe cluster density profiles, it has little theoretical justification and is not a very good fit to the inner regions, particularly for the dark matter. A better fit to the dark matter profile for $r < r_{200}$ is given by the simple fitting function

$$\rho/\overline{\rho} = \frac{1500\ r_{200}^3}{r\ (5\ r + r_{200})^2}. \qquad (8)$$

This function behaves as $r^{-1}$ at small radius in agreement with our best resolved models and with the much higher resolution simulations of Dubinski and Carlberg (1991).

The thick dotted lines in Figure 6b show Bertschinger's (1985) self-similar infall solution. This describes spherical accretion of collisionless and collisional fluids onto a point mass in an otherwise unperturbed Einstein-de Sitter universe. It predicts $\rho \propto r^{-9/4}$ at small radius. The simulated profiles are somewhat shallower than the similarity solution, particularly for the dark matter. This is not completely surprising, for profiles shallower than $\rho \propto r^{-9/4}$ have already been found in several analytic and numerical works (Fillmore and Goldreich 1984, Quinn *et al.* 1986, Hoffman 1988, and Efstathiou *et al.* 1988). The discrepancy is most marked near the centre where the simulated profiles fall well below Bertschinger's solution. This appears to be a genuine disagreement since it persists to radii several times the softening length in our best resolved clusters. However, still higher resolution simulations are needed for a definitive determination of central cluster profiles.

The scaled density profiles of the largest progenitors of the $CL1$-$CL6$ clusters at $z = 1$ are shown in Figure 7. Beyond the region dominated by the gravitational softening, these profiles are very similar to those at $z = 0$, which are illustrated by the thick solid lines. (Note that the slopes are the same in these regions; the small offset is due to our definition of $r_{200}$, which forces all the curves to encompass the same mass within this radius.) Near the center, the similarity breaks down, due to the smaller number of particles per cluster (up to a factor of ten), and to the fact that the gravitational softening is now a much larger fraction of $r_{200}$. As a result, the central overdensities are on average about three times lower than at $z = 0$. This can lead to systematic underestimation of physical quantities that depend crucially on the central properties of the clusters, such as the X-ray luminosity (see §3.6 below).

Although the gas and dark matter profiles appear very similar in Figure 6, the gas is actually slightly more extended than the dark matter. Inside $r_{200}$ the gas fraction $\Omega_{gas}$ is always smaller than the global value of 0.1 (see Table 1). This difference is also apparent in the ratio of the half mass radii of the gas and dark matter, $R_{gas}/R_{DM}$, which, as noted in Table 1, is typically greater than unity. (Note that these radii are determined relative to the total mass within $r_{200}$.) This property reinforces the conclusion of White *et al.* (1993) that the large gas fractions observed in real clusters cannot be reconciled with the much smaller global fraction expected from Big Bang nucleosynthesis considerations in an $\Omega = 1$ universe.

The segregation between dark matter and gas results from the different behaviour of the two components during a merger. Shocks prevent colliding gas distributions from interpenetrating. This produces a transient offset between gas and dark matter and a net transfer of energy and angular momentum to the gas (Navarro and White 1993). The effect can be illustrated by following the mean properties of the particles that end up in the core of a cluster, defined to be the region containing the inner 10% of its mass. The trajectory of each cluster core from $z = 1$ to $z = 0$ is shown in the density-temperature or density-velocity dispersion plane in Figure 8. The upper dotted line shows the track corresponding to isentropic gas, $\rho_{gas} \propto T^{3/2}$, while the lower dotted line corresponds to constant phase-space density for the dark matter, $\rho_{DM} \propto \sigma^3$. Shocks increase the entropy of the core gas and cause temperatures to rise with relatively little change in density. In contrast, the dark matter evolves at roughly constant phase-space density. As a result, the gas density in the cluster core is only 0.07 times the dark matter density, whereas a ratio of $\Omega_g/(1 - \Omega_g) = 0.11$ would be expected if the two distributions were identical.

### 3.3    Temperature profiles

Figure 9 shows scaled gas temperature profiles for our six clusters. These were obtained by normalizing to the characteristic temperature, $T_{200} = \mu m_p GM_{200}/(2kr_{200})$, where $k$ is Boltzmann's constant. The profiles agree to within about a factor 2, a similar level to that seen earlier for the density profiles. (Note the very different scales in the ordinates of Figures 6 and 9.) All the clusters have a near-isothermal inner region that extends, in the best resolved case, for about a decade in radius. Near $r \sim 0.4r_{200}$, the temperature begins to drop and by $r_{200}$ it has fallen to about half the central value. This is as expected for gas in hydrostatic equilibrium given the density profiles of Figure 6. In the region $0.1 \lesssim r/r_{200} \lesssim 0.4$, the gas and dark matter density profiles have an effective slope of $\sim -2$ (indicated by the thick solid line in Figure 6a), whereas near $r/r_{200} \simeq 1$, the profiles steepen to $r^{-2.5}$. The equilibrium temperature is therefore nearly constant in the inner region and drops approximately as $r^{-1/2}$ in the outer region.

The thick dotted line in Figure 9 shows the gas temperature in Bertschinger's (1985) similarity solution. In this model the temperature increases inwards from the shock and approaches an $r^{-1/4}$ power-law near the centre. In the outer cluster this model gives an acceptable match to the simulations, but in the inner parts it overestimates the temperature just as it overestimates the density (see Figure 6).

The temperature profiles that we find agree with that of Evrard (1990b) but disagree with that of Thomas and Couchman (1992). (Since the average temperature of the cluster in each of these studies is $\sim 10^8$K, the corresponding virial radius is $r_{200} \simeq 3$ Mpc.) In Evrard's calculation the near-isothermal region extends almost to $r_{200}$, as is the case in one of our own models, whereas the temperature drops by almost an order of magnitude between $0.1r_{200}$ and $r_{200}$ in Thomas and Couchman's cluster. None of our clusters shows such a dramatic decline in temperature. The reason for this discrepancy is unclear. All three studies assume the same cosmology and attempt to model the same physics using similar codes. Both Evrard and Thomas & Couchman select clusters as high peaks of the initial linear density field while we select clusters from the final nonlinear mass distribution. This seems unlikely to be the source of the problem given the dramatic difference between the two earlier simulations. It is possible that Thomas and Couchman's cluster is simply an atypical example.

### 3.4    Hydrostatic equilibrium and binding mass estimates

Using the density and temperature profiles derived in the preceeding subsections we can quantify how well hydrostatic equilibrium describes the intracluster medium of our simulated clusters. For real clusters the interpretation of X-ray data, and in particular the estimation of cluster mass, is usually based on the assumption that the gas is in equilibrium. Our simulations allow us to determine the accuracy of such estimates and to assess the biases which may arise when only incomplete information is available.

For a spherically symmetric cluster with the density profile of eqn (7), the equation of hydrostatic equilibrium may be written as

$$M_T(r) = \frac{k}{G\mu m_p} \left( \frac{3\beta_f (r/r_c)^2}{1 + (r/r_c)^2} + \frac{d\ln T}{d\ln r} \right) T(r)r. \qquad (9)$$

In Figure 10 we compare the mass derived from this equation with the actual mass, $M_{tot}$, contained within radius $r$. Outside the region where gravitational softening is important, eqn (9) underestimates the true mass by up to $\sim 30\%$. As pointed out by Evrard (1990b), this discrepancy reflects the existence of bulk motions in the gas whose kinetic pressure makes a non-negligible contribution to the total pressure support.

The importance of kinetic pressure in the simulations may be quantified by the parameter $\beta_{gas} = \mu m_p \sigma_{gas}^2/kT$, where $\sigma_{gas}$ is the velocity dispersion of the gas particles. For hydrostatic equilibrium, $\beta_{gas} = 0$. The values of $\beta_{gas}$ for our simulated clusters (Table 3) range between $\sim 0.1$-0.3 and are large enough to account for the discrepancy between the true mass and that estimated from eqn (9) and for the fact that the measured temperatures in Figure 9 fall slightly below the "virial temperature." The magnitude of the residual bulk motions depends on the detailed dynamical history of the system so a constant "fudge factor" for the mass estimates, as proposed by Evrard (1990b), is a rather crude approximation. Systems that have not experienced recent mergers (eg the solid curve in Figure 10) satisfy eqn (9) very well at all radii.

Although the ASCA X-ray satellite is now beginning to measure gas temperature profiles in clusters, for most objects only a single emission-weighted temperature is currently available. In this case, the gas is commonly assumed to be isothermal and the logarithmic derivative of the temperature in eqn (9) set equal to zero. The resulting mass estimates are shown in the lower panel of Figure 10. In most cases, the true mass is underestimated by a similar factor to that obtained when the temperature profile is known, reflecting the fact that the temperature distribution of our clusters is close to isothermal in the region where most of the X-ray emission is produced.

## 3.5   Scaling Laws

We now examine how well our simulated clusters obey the scaling relations described in the Introduction. The filled circles in Figure 11 show correlations between X-ray and dynamical properties for clusters $CL1$-$CL6$. (The filled triangles correspond to our "preheated" gas simulations and the crosses to data for real clusters; these will be discussed below.) Here, we consider the properties of the clusters at $z = 0$, rather than those of the "equilibrium" configurations discussed in the three preceeding subsections. Numerical values of the quantities plotted are given in Table 1.

The solid lines in Figure 11 show the scaling relations derived from eqns (1) and (2). The mass-temperature relation is normalized assuming that the mean overdensity of the clusters is 200: $T = 5.1(M_{200}/10^{15}\,\mathrm{M_\odot})^{2/3}$ keV. The mass-velocity dispersion is normalized assuming that the clusters are isotropic isothermal spheres: $\sigma_{DM} = (0.5GM_{200}/r_{200})^{1/2}$ km/s. The calculation of the X-ray luminosity requires additional assumptions about the structure of the clusters (e.g. the size of the core radius), so we fix the normalization of the mass-luminosity relation empirically by fitting to the results of the simulations. This gives $L_X = 1.2 \times 10^{45}(M_{200}/10^{15}\,\mathrm{M_\odot})^{4/3}$ erg/s. The normalizations of the other scaling relations follow from these. Overall, our simulations follow the scaling relations remarkably well. There is a small offset in the $T$-$M_{200}$ relation which results from the fact that the gas is incompletely thermalized and thus not in strict hydrostatic equilibrium. This also affects the $T$-$\sigma_{DM}$ relation.

## 3.6   Sensitivity to numerical parameters

The resolution of our simulations is determined primarily by the total number of particles, $N$, which effectively sets the minimum value of the gravitational softening, $h_g$, required to suppress two-body relaxation. Figure 12 shows the effect of increasing and decreasing the number of particles in the $CL1$ cluster by factors of 3 and 2 respectively. The largest simulation (dotted lines, run $CL1 \times 3$) has total $N \simeq 70,000$ and the smallest one (dashed lines, run $CL1/2$) has $N \simeq 11,000$. The solid lines correspond to the original $CL1$ simulation. Even though the total number of particles varies by a factor 6, the density and temperature profiles are practically indistinguishable.

As shown in Tables 1 and 3, dynamical properties are also in good agreement in the three runs ($CL1$, $CL1 \times 3$ and $CL1/2$). For example, the residual bulk motions of the gas ($\beta_{gas}$ in Table 3) are fairly insensitive to $N$ and thus seem to be unaffected by discreteness effects in the evolution of the cluster potential. Even the X-ray luminosity, the quantity most sensitive to numerical resolution because of its strong dependence on central density, seems to have converged. Increasing $N$ by a factor 3 increases $L_X$ by less than 10% although decreasing $N$ by a factor of 2 causes $L_X$ to drop by $\sim 30\%$. Decreasing the number of particles even further results in severe underestimates of the total X-ray luminosity of the cluster. This is shown in Figure 13, where we plot how $L_X$ depends on the total number of gas particles inside $r_{200}$ ($N_{gas}$). Only when $N_{gas}$ is larger than about 2,000-3,000 do the X-ray luminosity estimates seem to be reliable.

Other factors that may influence the central regions of a cluster, and therefore might spuriously affect the X-ray luminosity of a simulated cluster, include the choices of gravitational softening, initial redshift, and SPH smoothing lengths. We have carried out a number of experiments to test these possibilities. Varying $h_g$ from 100 to $200kpc$, $1+z_i$ from 4.74 to 9.48, and requiring 100 neighbours instead of the usual 32 in the SPH smoothing algorithm cause, however, little change in the X-ray luminosities at $z = 0$ for simulations with our standard particle number. Therefore, we conclude that the density profiles of these clusters have a well-defined "core" that is *resolved* in our simulations at $z = 0$. (We use the term "core" here rather loosely to signify the radius where the total X-ray luminosity starts to converge.) Inside this "core" the density is *not* constant, but its logarithmic slope is relatively shallow, so that once the numerical resolution is good enough, further improvements have no appreciable effect on $L_X$. Note that were this "core" produced by the gravitational softening we would not recover the $L_X \propto T^2$ relation predicted by the scaling laws, since $h_g$ does not scale from cluster to cluster. The gravitational softening was kept constant in *physical* units, and so the softened region is a different fraction of the virial radius in different clusters (cf. Figure 6).

The effects of limited numerical resolution can have dramatic effects on the evolution of the X-ray luminosity. At early times clusters are less massive and therefore more poorly resolved than at late times. This problem is compounded by the fact that at early times the gravitational softening, which is fixed in physical units, is a larger fraction of the virial radius. As shown by the $z = 1$ density profiles in Figure 7, these two effects significantly limit the central density that a cluster can attain. At this redshift our clusters contain typically no more than $\sim 400$ particles inside their virial radii. Figure 13 shows that this must severely compromise numerical estimates of their X-ray luminosity of a cluster, leading to a systematic underestimation of $L_X$ at high redshift and to a breakdown of the scaling behaviour. An effect of this magnitude is expected due to lack of resolution at this redshift (see Figure 13). This is illustrated in Figure 11 where we plot the $L_X$-$M_{200}$ relation for the progenitors of clusters $CL1$-$CL6$ at $z = 1$ (open circles). The dashed line shows the expected behaviour predicted by eqn (2). The redshift evolution of the simulated clusters is much weaker than the $(1 + z)^{7/2}$ dependence implied by the scaling laws; they are about 4 times less luminous than predicted. An effect of this magnitude can be entirely due to lack of resolution at this redshift, as shown in Figure 13. Although based on the present set of simulations we cannot *demonstrate* that the redshift evolution agrees with the theoretical expectations, the similarity of the "scaled" clusters in the regions not affected by numerical limitations leads us to suggest that the profiles shown in Figure 6 are generic for clusters formed in this model, and that the departure of simulated clusters from the predicted evolutionary behaviour is likely to be due mainly to resolution effects.

The dependence of the estimated X-ray luminosity on numerical resolution is not surprising but has sometimes been overlooked. Kang *et al.* (1994) and Bryan *et al.* (1994) interpreted a discrepancy between the redshift evolution of the X-ray luminosity and the scaling laws as a genuine pre-diction of their models rather than as a reflection of their numerical limitations. For the same reason, their conclusions regarding clusters at the present epoch may also be affected by resolution effects. Since their clusters were identified in single simulations of cosmological volumes, they are resolved to varying degrees. Poorer clusters have smaller numbers of particles and subtend fewer resolution lengths. As a result their central densities and X-ray luminosities may be systematically underestimated by a larger factor than those of richer clusters. This would lead to a steeper $L_X$-$T$ relation than the scaling laws predict. At $z = 0$, our simulations are not affected in this way because we use approximately the same number of particles to simulate clusters of differing mass. Our derived $L_X$-$T$ relation thus agrees remarkably well with the theoretical scaling relation.

## 4   COMPARISON WITH REAL CLUSTERS

### 4.1   Gas density profiles

The gas density profiles discussed in Section 4.2 are moderately well fit by the "$\beta$-models" (eqn 7) often used to describe observed clusters. Since the core regions of real clusters are affected by cooling processes which we have neglected, we limit our comparison to the outer profile, as measured by the parameter $\beta_f$. As noted by Evrard (1990a), typical values of $\beta_f$ found in simulations of this kind are $\sim 0.8$, rather than $\sim 0.6$, the value inferred for most observed clusters from their X-ray surface brightness profiles. However, this comparison can be misleading. Since eqn (7) is not a perfect description of our models, the "best" value of $\beta_f$ depends on the range of radii used for the fit. We illustrate this in Figure 14 by plotting the value of $\beta_f$ derived for the ensemble of Figure 6b against $r_{max}$, the outermost radius used in the fit. While we find $\beta_f \sim 0.8$ for $r_{max} \simeq r_{200}$, $\beta_f$ is only $\sim 0.6$ for $r_{max} \simeq 0.2 r_{200}$

The value of $r_{max}$ used in analyses of real clusters depends on the quality of the data. The filled circles in Figure 14 show values of $\beta_f$ for those clusters from the compilation of Jones and Forman (1984) for which we have been able to infer the value of $r_{max}$. We assume $r_{max}$ to be the radius at which uncertainties in the measured surface brightness exceed a factor of 10 (see Figures 1 and 2 of Jones and Forman; we take $r_{200} \approx 3(T_X/7.5\text{keV})^{1/2}$ Mpc for their clusters). The point at $r_{max}/r_{200} = 1$ corresponds to the ROSAT observations of the Coma cluster reported by Briel *et al.* (1992). The trend in the real data is similar to that found in the simulations. In most cases, the observations become noisy well inside the virial radius and this may account for the low $\beta_f$ values measured. As discussed by Evrard (1990a), a further bias could result from incorrect background subtraction. The $\beta_f$ values for real clusters lie slightly below those for the simulations but overall there is reasonable agreement. Higher signal-to-noise observations should help to clarify these issues.

### 4.2   Temperature profiles

The temperature profiles of our clusters are characterized by a near-isothermal region extending to $r \sim 0.5 r_{200}$ and a gradual decline to about half the central value near $r_{200}$. Profiles for many real clusters should soon be available from observations with the ASCA satellite, but for the time being we compare our results with the Coma cluster for which a radial temperature profile has been measured to $\sim 1$ Mpc (Watt *et al.* 1992), about $1/3$ of the virial radius. In this region, the cluster is very close to isothermal. Although the temperature further out has not been directly measured, detailed modelling of X-ray and optical data by Hughes *et al.* (1988a,b) suggests a drop in temperature similar to that seen in our models. Hughes *et al.* derive a polytropic index, $\Gamma \simeq 1.6$ for the outer parts of Coma. Values of $\Gamma$ obtained by fitting a power law to the temperature profiles of our clusters in the range $0.4 < r/r_{200} < 1$ (see Figure 9) are listed in Table 2. These values are somewhat lower than the value quoted by Hughes *et al.* but, given the considerable uncertainty in the analysis of the data, we conclude that, as with the density profiles, the temperature profiles of our simulated clusters are consistent with those inferred for real clusters.

### 4.3   Correlations

X-ray luminosities, gas temperatures and galaxy velocity dispersions for various samples of clusters are plotted in the correlation diagrams of Figure 11. The crosses represent data from Edge and Stewart (1991a,b), Henry and Arnaud (1991), David *et al.* (1993), and Lubin and Bahcall (1993). Starred polygons correspond to two Hickson groups (HG92 and HG62) recently detected with ROSAT (Mulchaey *et al.* 1993, Ponman and Bertram 1993).

The fundamental $L_X$-$T$ relation for the data obeys the well known law, $L_X \gtrsim T^3$, with the two Hickson groups falling on the extrapolation of the relation for rich clusters. This is significantly steeper than the corresponding relation for our simulated clusters which closely follow the scaling law of eqn (2), $L_X \propto T^2$. Thus, at low temperatures, our models give about an order of magnitude higher X-ray luminosity than is observed. The discrepancy between our models and the observed $L_X$-$T$ relation is reflected in the $L_X$-$\sigma$ plane although in this case the scatter in the data is larger. By contrast, the observed $T$-$\sigma$ relation does not show any systematic deviation from the relation obeyed by our models, $T \propto \sigma^2$, appropriate to an isothermal gas in hydrostatic equilibrium. However, the scatter in the data is larger than in the models, a feature related to the "$\beta$-discrepancy" which we discuss in some detail below.

Several suggestions have been put forward to explain why the simple nonradiative models fail to account for the observed $L_X$-$T$ relation. One possibility is that the gas fraction increases with cluster temperature (David *et al.* 1990, Arnaud *et al.* 1992). Another is that cooling preferentially depresses the luminosity of poorer clusters. A third (which we discuss further below) is that the central gas densities and core radii vary systematically with temperature, as expected if the intracluster medium were heated to relatively high entropy at early times.

### 4.4   The "$\beta$-discrepancy"

The "$\beta$-discrepancy" in clusters is a problem which has exercised X-ray astronomers for many years. It refers to a discrepancy which arises when attempting to fit cluster data to the $\beta$-model proposed by Cavaliere and Fusco-Femiano (1976). This model assumes that clusters are spherical, are in hydrostatic equilibrium, and that both the intracluster medium and the galaxies may be modelled as isotropic isothermal "gases" with density profiles given by eqn (7). With these assumptions the hydrostatic equilibrium equation can be integrated immediately to give the gravitational potential of a cluster,

$$\Phi(r) = \frac{3\beta_f}{2} \frac{kT}{\mu m_p} \ln(1 + r^2/r_c^2). \tag{10}$$

**Table 3.** Values of the $\beta_T$ parameter.

| Model | $\beta_T(T_X)$ $[r_{200}]$ | $\beta_T(T_m)$ $[r_{200}]$ | $\beta_{gas}(T_X)$ $[r_{200}]$ | $\beta_T(T_X)$ $[r_{1000}]$ | $\beta_T(T_m)$ $[r_{1000}]$ | $\beta_{gas}(T_X)$ $[r_{1000}]$ | $\beta_T(T_X)$ $(z=0)$ |
|---|---|---|---|---|---|---|---|
| $CL1$ | 1.01 | 1.18 | 0.11 | 1.08 | 1.10 | 0.07 | 1.01 |
| $CL2$ | 1.07 | 1.14 | 0.15 | 1.15 | 1.17 | 0.11 | 1.28 |
| $CL3$ | 1.18 | 1.36 | 0.36 | 1.31 | 1.32 | 0.37 | 1.37 |
| $CL4$ | 1.04 | 1.18 | 0.22 | 1.08 | 1.18 | 0.12 | 1.24 |
| $CL5$ | 1.05 | 1.23 | 0.08 | 1.24 | 1.16 | 0.06 | 1.18 |
| $CL6$ | 1.09 | 1.38 | 0.30 | 1.13 | 1.10 | 0.19 | 1.09 |
| $CL1 \times 3$ | 0.94 | 1.17 | 0.10 | 1.05 | 1.11 | 0.06 | 0.94 |
| $CL1/2$ | 0.92 | 1.12 | 0.09 | 1.00 | 1.06 | 0.07 | 0.92 |

Note that the density corresponding to this potential falls as $r^{-2}$ at large radii and that it *cannot* be cast in the form of equation (7). Since gas and galaxies are in equilibrium in the same potential well, equation (10) shows that the two components must have the same value of the product $\beta_f T$. Thus if we retain $\beta_f$ and $T$ for the properties of the gas, and denote the corresponding properties of the galaxy distribution by $\beta_{f*}$ and $T_* = \mu m_p \sigma_*^2/k$ where $\sigma_*$ is the one-dimensional velocity dispersion, the slopes of the density profiles of the two components are related to the ratio of their temperatures by

$$\beta_f/\beta_{f*} = \frac{T_*}{T} = \frac{\mu m_p \sigma_*^2}{kT} \equiv \beta_T. \tag{11}$$

The galaxy distribution is often modelled by King's approximation to the core of an isothermal sphere (eqn 7 with $\beta_{f*} = 1$). In this case, the slope of the gas density profile, $\beta_f$, should equal $\beta_T$. As we saw earlier, fits to X-ray surface brightness profiles typically give $\beta_f \simeq 0.6$ (Bahcall and Sarazin 1977, Mushotzky 1984, Jones and Forman 1984), whereas measurements of gas temperatures and galaxy velocity dispersions give $\beta_T \simeq 1$. (Many values of $\beta_T$ which were found to be significantly greater than one in early studies have subsequently been revised downwards as a result of improved velocity dispersion determinations. Edge and Stewart (1991a,b) obtained $\langle \beta_T \rangle = 0.91 \pm 0.08$ for 23 clusters, almost identical to the value obtained by Bahcall and Lubin (1994) from a larger compilation.) Clearly, either one or more of the assumptions of the above model is invalid or some of the measurements are in error.

If we assume the "galaxies" to be distributed in the same way as the dark matter in our simulations, and we analyze them in the same way as the real data, we find a similar "$\beta$-discrepancy". When we fit the gas density profiles in the inner parts only, we obtain, $\beta_f \simeq 0.6$ (Figure 14), while a comparison of the velocity dispersion of the dark matter with the *emission-weighted* X-ray temperature yields $\beta_T \simeq 1$ (column 2 of Table 3). This apparent discrepancy is readily explained as follows: (*i*) the model gas density profiles become steeper with radius so that if eqn (7) is fitted to the entire radial range (not just to the inner parts), the mean value of $\beta_f$ for our six clusters is $\beta_f = 0.82 \pm 0.06$, not $\beta_f = 0.6$ (Table 2); (*ii*) the dark matter density profile is shallower than King's approximation and has $\langle \beta_{f*} \rangle = 0.86 \pm 0.11$ (Table 2) rather than $\beta_{f*} = 1$. Hence $\beta_f/\beta_{f*} \approx \beta_T$ as expected.

It is quite likely that a similar resolution of the "$\beta$-discrepancy" applies to real clusters. As we showed in Section 4.1, the available data suggest that their gas density profiles differ from perfect $\beta$-models in the same way as our simulations do. In addition, as Bahcall and Lubin (1994) have emphasized, the galaxy density profile in real clusters is better fit by $\beta_{f*} \simeq 0.8$ than by $\beta_{f*} \simeq 1$. It is clear that since the gas and galaxy distributions in clusters are neither isothermal nor perfect fits to equation (7), an analysis in terms of these models must be treated with some caution.

Our results are similar to those of Evrard (1990a,b), but our interpretation is slightly different. Evrard's clusters also exhibit a $\beta$-discrepancy, but the values of $\beta_f$ and $\beta_T$ he derives are significantly larger than ours. His value of $\beta_T \simeq 1.2$ exceeds ours because he used a mass-weighted temperature whereas we use an emission-weighted temperature. Since the models are not isothermal, the two weightings give different results. The emission-weighted temperature is closer to the quantity measured for real clusters. In Table 3 we compare values of $\beta_T$ using emission-weighted ($T_X$) and mass-weighted ($T_m$) temperatures for the clusters as a whole ($r_{200}$) and for the inner regions alone ($r_{1000}$). (All columns in this table except the last one refer to the "equilibrium" configurations discussed in Section 4.2; $r_{1000}$ is the radius of a sphere of mean overdensity 1000.) Evrard ascribed the $\beta$-discrepancy in his models to a combination of incomplete thermalization of the gas and to anisotropies in the velocity distribution of the dark matter. Incomplete thermalization is evident in our models also, but, as Table 3 shows, it has a relatively weak effect on the estimate of $\beta_T$ when emission-weighted rather than mass-weighted temperatures are used. Once these effects are taken into account, our results are in good agreement with Evrard's and with those of Thomas and Couchman (1992).

The largest values of $\beta_T$ occur in clusters which have undergone recent mergers. Mergers stir up the gas in the outer parts and the resulting departure from hydrostatic equilibrium can give rise to values of $\beta_T$ as large as $\sim 1.4$. This is illustrated in the last column of Table 3 which gives data for clusters identified at $z = 0$ (within $r_{200}$), rather than for the "equilibrium" configurations discussed above. At $z = 0$ only $CL1$ and $CL6$ are undisturbed and these have $\beta_T(T_X) \simeq 1$. Provided the measured galaxy velocity dispersion is reliable, values of $\beta_T$ significantly in excess of one can be taken as an indicator of recent merger activity.

None of our clusters has a $\beta_T$ value significantly below unity, yet there are several clusters in the samples of Edge and Stewart (1991a,b) and Lubin and Bahcall (1993), particularly low-temperature clusters, with $\beta_T < 1$. Unless the temperature of these clusters has been overestimated, such low values of $\beta_T$ indicate a bias in the galaxy velocity dispersion similar to that seen in the simulations of Carlberg and Couchman (1989), Carlberg and Dubinski (1991), Couchman and Carlberg (1992), Katz *et al.* (1992) and Evrard *et al.* (1994a). Such "velocity bias" should in principle be observable in the form of a steep number-density profile for the galaxies (*i.e.* in a large value of $\beta_{f*}$).

## 5    "PREHEATED" GAS

We have seen that a simple model in which the intracluster medium originates from an initially cold, nonradiative gas leads to density and temperature profiles which resemble those observed in real clusters. Nevertheless, the X-ray luminosities of our clusters do not match observations very well. In particular, the predicted correlation between X-ray luminosity and temperature is too shallow ($L_X \propto T^2$ rather than $L_X \gtrsim T^3$) and, for realistic cluster mass functions, the amplitude of the X-ray luminosity function increases with redshift, in conflict with observation. Since most of the X-ray luminosity originates in the cluster core, it seems worthwhile to investigate variants of the simple model which modify the state of the gas in the inner regions. One such variant, originally proposed by Kaiser (1991) and Evrard and Henry (1991), replaces the assumption that the gas is initially cold with the assumption that it was heated prior to cluster collapse by non-gravitational processes associated with galaxy formation. Placed on a high initial adiabat, the gas in the central regions is less affected by shocks and evolves at nearly constant entropy. Eqns (1) and (3) give the scaling laws describing a cluster population in which all clusters are assumed to be isothermal and to have the same central specific entropy.

To investigate this model in a quantitative fashion, we repeated the $CL1$, $CL3$ and $CL5$ simulations, this time setting the initial temperature of each gas particle so that its specific entropy was about the same as the final value attained in the core of $CL1$, our most massive cluster (see Figure 4d). The resulting three simulations are labelled $EP1$, $EP2$ and $EP3$ (see Table 1). The right-hand panel of Figure 8 shows that the core regions of the preheated clusters do indeed evolve isentropically. As a result, the central gas density no longer exhibits the scaling behaviour of our earlier models. This is illustrated in Figure 15 where the thick solid line shows the ensemble fit to the density profile of the (initially cold) clusters of Figure 6b. The central overdensities of the new models are lower and depend on the mass (or temperature) of the cluster. Cooler clusters have lower central overdensities than hotter clusters. Note, however, that changes due to the preheating are seen only in the central regions. In the outer cluster the density profiles are almost unchanged.

Since the effects of preheating are greater in cooler clusters, X-ray luminosity increases more rapidly with mass and temperature in this model. This is shown in the two central panels of Figure 11 where the $EP$ clusters are represented by solid triangles. The dotted lines indicate the prediction of eqn (3) for evolution at constant central entropy. The simulations agree extremely well with this prediction and give a much better fit to the observed $L_X$-$T$ correlation (crosses and stars) than the initially cold clusters. Because of the numerical limitations discussed in Section 3.6, we cannot reliably determine the evolution of the X-ray luminosity function. The scaling relations obeyed by our clusters at $z = 0$ imply that the $L_X$-$T$ correlation should be independent of redshift and this, in turn, predicts a decline in the abundance of high luminosity clusters with increasing redshift. The rate of evolution depends on the initial fluctuation spectrum and for plausible choices it appears to be comparable to that seen in the limited datasets available (Evrard and Henry 1991, Castander *et al.* 1994). Finally, we note that the anticorrelation between the core radius (in units of $r_{200}$) and mass seen in Figure 15 implies that the core radius of the gas distribution (in physical units) is essentially independent of mass. This agrees with the scaling, $r_{core} \propto M^{-1/6}$, expected if all clusters have the same central entropy (cf. eqns 1 and 3).

## 6    CONCLUSIONS

Our simulations of cluster formation make similar physical assumptions and employ similar numerical techniques to those of Evrard (1990a,b). Nevertheless there are some significant differences; the present experiments have better resolution, follow clusters with a wider range of mass, and choose clusters according to the final nonlinear mass distribution rather than the initial linear overdensity field. Although our results agree well with Evrard's earlier work, we have chosen to address somewhat different issues. In particular, we have investigated whether, in the absence of cooling or heating, the distributions of mass and of gas can be treated as varying from cluster to cluster only through a scaling with size and age. We have also explored a model in which the gas is assumed to be heated before cluster collapse. Our main conclusions are the following.

1) In a flat universe cluster formation by hierarchical clustering does indeed lead to objects whose properties scale at least approximately with size and age. Measuring radii in units of the radius enclosing a mean mass overdensity of 200, densities in units of the cosmic mean, and temperatures in units of the corresponding virial value, we find that (to within a factor of 2 and over the limited radius range modelled reliably in our simulations) the density and temperature profiles have a shape which depends at most weakly on time and on cluster mass. This result is expected, since the CDM power spectrum is virtually scale-free over the relevant range of length-scales and, in the absence of heating and cooling processes, there are no characteristic physical scales in the problem. Our success in verifying this scaling over a wide range of physical and numerical parameters suggests that our techniques are adequate to resolve the structure of present-day clusters.

2) Gas and dark matter evolve differently in the cores of clusters. Shocks cause a continual increase of gas entropy, while the maximum phase space density of the dark matter remains approximately constant. Numerical resolution is a significant limit on our ability to study dark matter cores but may not affect our estimates of the central thermodynamic properties of the gas. Although simulations with better resolution are required to confirm this conjecture, we note that Pearce *et al.* (1994) come to very similar conclusions on the basis of their more schematic but higher resolution simulations of cluster merging.

3) Our equilibrium clusters have X-ray properties which obey the standard scaling laws (*e.g.* Kaiser 1986). In particular, at $z = 0$ our clusters satisfy $L_X \propto T^2$ very well, and we find that clusters of a given mass are more luminous at higher redshift. Unfortunately, our simulations have insufficient resolution for a proper test of the scaling of luminosity with redshift. Previous studies using Eulerian hydrodynamic techniques (e.g. Kang *et al.* 1994, Bryan *et al.* 1994) are likely to have predicted incorrect evolutionary behaviour and an incorrect $L_X - T$ relation because of poor resolution in the central regions of clusters.

4) The dark matter and the gas have similar density profiles in our equilibrium clusters, but the gas is slightly more extended. The slope $\alpha = d \ln \rho / d \ln r$ of the gas density profile increases gradually from $\alpha \sim -1$ at $r \sim 0.05 r_{200}$ to $\alpha \sim -2.5$ near $r_{200}$. A standard $\beta$-model yields $\beta_f \sim 0.8$ if the whole range $0.03 < r/r_{200} < 1$ is used, but lower values are found if, as is common in observational studies, only the central regions are fitted. The gas temperature is almost constant in the central regions but tends to drop beyond $\sim 0.4 r_{200}$. At $r_{200}$ it is typically half of the central value. These density and temperature profiles agree reasonably well with those inferred for observed clusters without strong cooling flows.

5) The parameter $\beta_T = \mu m_p \sigma_{DM}^2 / kT$ is close to unity for our equilibrium clusters provided emission-weighted temperatures are used. Larger values are often associated with ongoing mergers. None of our simulated clusters has $\beta_T$ much less than 1, suggesting that "velocity bias" (*i.e.* a lower velocity dispersion for the galaxies than for the dark matter) may be substantial in those clusters observed to have $\beta < 1$. In our models, as in real data, $\beta_T > \beta_f$. As has been frequently noted (*e.g.* Sarazin 1988, Evrard 1990b, Bahcall and Lubin 1994) this infamous "$\beta$-discrepancy" can be traced to the fact that the galaxy and dark matter distributions are not well represented by King's approximation to the core of an isothermal sphere.

6) Although formation models with a nonradiative and initially cold gas match most observed properties of clusters, their core structure results in X-ray luminosities which obey neither the observed scaling with temperature nor the observed evolutionary trend. As pointed out by Kaiser (1991) and Evrard and Henry (1991), models in which the gas initially has high entropy are in much better agreement with observation. Such preheating only affects the core properties of the gas; the outer density profiles are essentially unchanged. However, since observed cluster cores often have short cooling times, it is difficult to imagine that their specific entropy is constant and reflects processes occurring prior to cluster formation. It is unlikely that a full explanation of the X-ray properties of clusters will emerge before the role of cooling in cluster cores is properly understood.

## ACKNOWLEDGEMENTS

We thank the referee, Peter Thomas, for a very careful reading of our paper. This work has been supported by the UK Science and Engineering Research Council.

**Figure 1:** Evolution of the distribution of gas (left) and dark matter (right) for cluster $CL1$. This is the most massive cluster in our ensemble of simulations. The epochs shown correspond to redshifts 2, 1, 0.4, and 0. Units are in *physical* kpc for $H_0 = 50$ km s$^{-1}$ Mpc$^{-1}$. Note the resemblance between the distributions of gas and dark matter at all times.

**Figure 2:** The gas distribution at the final time ($z = 0$) in clusters from the $CL$ series. The dark matter distribution (not shown) is similar to that of the gas in all cases. Units are in *physical* kpc for $H_0 = 50$ km s$^{-1}$ Mpc$^{-1}$. Note that 4 out of the 6 $CL$ clusters are undergoing a major merger event at $z = 0$.

**Figure 3:** The evolution of the dark mass inside a sphere of current mean overdensity 200 centred on the most massive clump. Masses are in units of the overdensity 200 mass at $z = 0$. Most of the cluster mass is accreted during merger events, reflected in sudden jumps in mass. Clusters form late; three of them assemble half of their mass only after $z = 0.2$.

**Figure 4:** Evolution of the hydrodynamic properties of the intracluster gas in the six $CL$ clusters. The "density" and "temperature" panels show the time evolution of the core (solid line), main body (dotted line), and outer boundary (dashed line) of the cluster. These properties were obtained by averaging over 10% of the particles which are nearest the core, half-mass radius and outer boundary respectively. The gas density and entropy are in code units (§2.3). For clarity, the temperature curves have been shifted downwards successively by one order of magnitude. The X-ray luminosity includes the emission due to all particles which at $z = 0$ make up each cluster. The "entropy" panel shows the evolution of the core and boundary of each system only. The initial entropy of "pre-heated" clusters ($EP$ series) is shown by a horizontal dotted line. Units are given in §2.3.

**Figure 5:** Gas and dark matter density profiles for equilibrium configurations of $CL$ clusters. The profiles were constructed by averaging the density in spherical shells containing 50 particles each. The units of density are given in §2.3. The different curves show different clusters in order of decreasing mass.

**Figure 6:** Scaled density profiles of $CL$ clusters. The solid line segment in the upper panel has the slope, $\rho \propto r^{-2}$, of an isothermal sphere. Gravitational softenings are shown as small vertical lines with the same line type as the corresponding profile. The lower panel also shows the best ensemble fit using the isothermal $\beta$-model (thick solid line) and Bertschinger's (1985) self-similar solution (thick dotted line).

**Figure 7:** As Figure 6, but for clusters identified at $z = 1$. The thick curves are the same as those shown in Figure 6. Note that at this earlier time, the profiles deviate from the $z = 0$ ensemble fits at increasingly large radii because the number of particles in each cluster is smaller and the gravitational softening lengths (shown as vertical lines) become an increasingly large fraction of $r_{200}$. Outside the core, the profiles are very similar to those at $z = 0$.

**Figure 8:** Evolution of core particles (innermost 10% of the mass at $z = 0$) in the density-temperature (gas) and density-velocity dispersion (DM) plane. Dotted lines show loci of constant gas entropy, $T/\rho^{2/3}$, and constant dark matter phase-space density, $\rho/\sigma^3$. Left and right panels show the evolution between $z = 1$ (open symbols) and $z = 0$ (filled symbols) of $CL$ and $EP$ clusters, respectively. Units are arbitrary and relative values of the "entropy" increase upwards.

**Figure 9:** Scaled temperature profiles of the $CL$ clusters. The thick dotted line shows Bertschinger's (1985) self-similar solution. The thick solid segment indicates $T \propto r^{-1/2}$. Vertical segments show the gravitational softening. All clusters are nearly isothermal out to $r \sim 0.4 r_{200}$ and typical temperatures at the virial radius are about one-half of the values near the center. The units are described in the text.

**Figure 10:** The ratio of binding mass (estimated from eqn 9) to total mass as a function of radius. Incomplete thermalization leads to a systematic underestimate of the true total mass.

**Figure 11:** Correlations between the bolometric X-ray luminosity, total mass (at mean density contrast 200), 1-$D$ dark matter velocity dispersion, and X-ray emission-weighted temperature, for $CL$ clusters (filled circles) and pre-heated clusters (filled triangles). Observational data compiled from sources quoted in the text are plotted as small crosses and starred polygons. Solid lines show the scaling relations (eqns 1 and 2) appropriate to an initially cold gas; dotted lines show the corresponding relations (eqns 1 and 3) for the case when all clusters have the same initial minimum entropy. Open circles in the luminosity-mass panel represent the progenitors of $CL$ clusters at $z = 1$ and the dashed line shows the evolution expected from the scaling relations. The weaker evolution seen in the simulations reflects the fact that clusters identified at high redshift are more poorly resolved.

**Figure 12:** Scaled density and temperature profiles for simulations with the same initial conditions as $CL1$ but with 3 times the total number of particles (70000 particles in total; dotted line) and 1/2 the total number of particles (11000 particles in total; dashed line). The solid line shows the profiles for the $CL1$ cluster. No significant differences are noticeable.

**Figure 13:** The dependence of the X-ray luminosity of cluster $CL1$ at $z = 0$ on the number of particles used in the simulation. Luminosities are in erg/s, and $N_{gas}$ is the number of gas particles inside the virial radius of the cluster. In order of decreasing $N_{gas}$, the points correspond to runs $CL1 \times 3$, $CL1$, $CL1/2$, (see Table 1). The other two runs correspond to decreasing the number of particles of the $CL1$ run by factors of 3 and 6, respectively.

**Figure 14:** The dependence of the parameter $\beta_f$ (eqn 7) on the outermost radius, $r_{max}$, used in the fit. Open circles represent the simulated clusters. Filled circles correspond to the eight clusters from Jones and Forman (1984) for which $r_{max}$ can be estimated (see text and their Figures 1 and 2) and to the Coma cluster from the results of Briel *et al.* (1992). In both models and real data, lower values of $\beta_f$ are obtained for lower values of $r_{max}$.

**Figure 15:** As Figure 6, but for clusters with pre-heated gas ($EP$ in Table 1). The three clusters have the same initial central entropy. More massive clusters have higher central densities and smaller core radii (in units of their virial radii) than smaller clusters.